\documentstyle[preprint,revtex]{aps}
\begin{document}
\draft
\preprint{May 31, 1993}
\begin{title}
{\bf Toward a Unified Magnetic Phase Diagram\\
of the Cuprate Superconductors}
\end{title}
\bigskip
\author{\large Alexander Sokol$^{(1-3)}$ and David Pines$^{(1)}$}
\bigskip
\begin{instit}
{\rm
$^{(1)}$Department of Physics and
$^{(2)}$Materials Research Laboratory,\\
University of Illinois at Urbana-Champaign, Urbana, IL 61801-3080}
\end{instit}
\begin{instit}
{\rm
$^{(3)}$ L.D. Landau Institute for Theoretical Physics, Moscow, Russia}
\end{instit}

\begin{abstract}
We propose a unified magnetic phase diagram of cuprate
superconductors. A new feature of this phase diagram is
a broad intermediate doping region
of quantum-critical, $z\!=\!1$, behavior,
characterized by temperature independent $T_1T/T_{\rm 2G}$
and linear $T_1T$, where the spin waves are not completely absorbed
by the electron-hole continuum.
The spin gap in the moderately doped materials
is related to the suppression of the low-energy spectral weight
in the quantum disordered, $z\!=\!1$, regime.
The crossover to the $z\!=\!2$ regime, where
$T_1T/T_{\rm 2G}^2\!\simeq\!\mbox{const}$,
occurs only in the fully doped materials.
\end{abstract}
\narrowtext

\bigskip
\noindent
{\bf 1.Introduction}
\bigskip

Recent measurements \cite{Imai7,ImaiT1,TakigawaT2,ImaiT2}
of the spin-echo decay rate, $1/T_{\rm 2G}$,
for a number of cuprate oxides, taken together with earlier
measurements of spin-lattice relaxation rate $1/T_1$, provide
considerable insight into their low frequency spin dynamics.
In this communication we show how these measurements
may be combined with straightforward scaling arguments to obtain
a unified magnetic phase diagram for the Y- and La-based systems.

In the presence of strong
antiferromagnetic correlations at a wavevector ${\bf Q}$,
the main contribution to both $T_1^{-1}$ and $T_{\rm 2G}^{-1}$
for copper comes from small
$ {\bf \widetilde q}={\bf q}-{\bf Q} $, so that one may write
\cite{Pennington:Slichter}:
\begin{equation}
\frac{1}{T_1T} \sim
\int d^2{\bf \widetilde q}
\lim_{\omega\to0} \frac{\chi''({\bf \widetilde q},\omega)}{\omega},
\ \ \ \
\frac{1}{T_{\rm 2G}} \sim
\left[
\int d^2{\bf \widetilde q} \, \chi^2({\bf \widetilde q},0)
\right]^{1/2},
\label{T12}
\end{equation}
where $\chi({\bf \widetilde q},\omega)$ is the
electronic spin susceptibility near ${\bf Q}$.
On making use of quite straightforward scaling arguments,
when applicable, one may
substitute $\chi({\bf \widetilde q},\omega)=\xi^{2-\eta}
\hat\chi({\bf \widetilde q} \xi,\omega/\bar{\omega})$
into Eq.(\ref{T12})
($-\eta$ is the scaling dimension of the real space spin correlator)
and obtain:
\begin{equation}
\frac{1}{T_1}\sim T\xi^{-\eta}\bar{\omega}^{-1}, \ \ \ \
\frac{1}{T_{\rm 2G}}\sim \xi^{1-\eta}, \ \ \ \
\frac{T_1T}{T_{\rm 2G}} \sim \xi\bar{\omega}.
\end{equation}
where $\xi$ is the correlation length and
$\bar{\omega}$ an appropriate energy scale.

\bigskip
\noindent
{\bf 2. Applicable Scaling Regimes}
\bigskip

We consider first a clean quantum antiferromagnetic
insulator (referred to as ``insulator'' hereafter),
described by the $S\!=\!1/2$ Heisenberg model with the
exchange coupling $J$.
Because the spin stiffness, $\rho_s\simeq 0.18J$ \cite{Singh},
is small compared to $J$,
the {\em quantum-critical} (QC) scaling regime \cite{CHN},
where the only energy scale is set by temperature,
$\bar{\omega}\!\sim\!T$,
exists over a substantial temperature range
$2\rho_s\!\alt\!T\!\alt\!J$ \cite{Sachdev:Ye,Chubukov:Sachdev}.
The dynamical exponent, $z$, which relates the
characteristic energy and length scales according to
$ \bar{\omega} \!\sim\! \xi^{-z}$,
is $z\!=\!1$ as a consequence of Lorentz invariance at $T\!=\!0$,
reflected in the linear dispersion relation of the spin waves.
In this case,
$ T_1T/T_{\rm 2G}\!\sim\!\bar{\omega}\xi
\!\sim\!\xi^{1-z}\!\simeq\!\mbox{const}$.
Since $\xi\!\sim\!\bar{\omega}^{-1/z}
\!\sim\!T^{-1/z}\!\sim\!T^{-1}$,
one further obtains
$1/T_1\!\sim\!T\xi^{z-\eta}\!\sim\!T^\eta $
\cite{Chubukov:Sachdev} and
$1/T_{\rm 2G}\!\sim\!\xi^{1-\eta}
\!\sim\!T^{\eta-1}$; because
the critical exponent $\eta$ is negligible,
$1/T_1\!\simeq\!\mbox{const}$,
while $1/T_{\rm 2G}\!\sim\!T^{-1}$.

A second regime of interest
is the two-dimensional {\em renormalized classical} (RC)
regime, $T_N\alt T\alt 2\rho_s$, which is characterized
by an exponential increase of the correlation length and
relaxation rates. In the dynamical scaling theory of
Chakravarty, Halperin, and Nelson \cite{CHN},
$1/T_1\!\sim\!T^{3/2} \exp(2\pi\rho_s/T)$ and
$1/T_{\rm 2G}\!\sim\!T \exp(2\pi\rho_s/T)$.
The prefactors arising
from the $\log \xi $ corrections
lead to a power-law temperature dependence of
the ratio $T_1T/T_{\rm 2G} \!\sim\! T^{1/2}$, while
$z\!=\!1$ leads to the cancellation
of the leading (exponential) terms.

According to numerical calculations for the insulator
in the 2D $S\!=\!1/2$
Heisenberg model \cite{Makivic:Jarrell,Sokol:Gagliano:Bacci},
as long as $T\!\alt\!J$,
the damping, $\gamma_q$,
of the high energy ($\omega_q\!\agt\!c\xi^{-1}$)
spin wave excitations,
is small throughout the Brillouin zone;
hence, for both RC and QC regimes
those can be treated as good eigenstates of the model.
The dynamical susceptibility can then be well approximated as:
\begin{equation}
\chi({\bf q},\omega) = \phi_q \left(
\frac{1}{\omega - \omega_q + i \gamma_q} -
\frac{1}{\omega + \omega_q + i \gamma_q}
\right),
\label{chi}
\end{equation}
except near the origin, where
the dynamics is diffusive as a consequence of
total spin conservation, and near the Neel ordering vector,
${\bf Q}\!=\!(\pi/a,\pi/a)$, where it is relaxational;
for $\xi^{-1}\!\alt\!\widetilde q\!\alt\!a^{-1}$,
the expression (\ref{chi}) is valid in both QC and RC regimes,
where $ \phi_q \sim 1/q $ and $ \omega_q\!\simeq\!cq $.
According to Ref.\cite{Hayden},
the one-magnon neutron scattering intensity in the insulator
is indeed well described by Eq.(\ref{chi})
with $ \gamma_q\!\ll\!\omega_q $.

It has been conjectured in Ref.\cite{Chubukov:Sachdev}
that the small doping as well as
randomness related to it are not likely to
affect the universal scaling behavior at high temperatures.
Quite generally, one would expect
a departure from $z\!=\!1$ behavior
only when spin waves become overdamped
by the electron-hole continuum \cite{Millis:z=2}.
Since this would require a substantial increase
in spin wave damping,
to $ \gamma_q\!>\!\omega_q$,
compared to its insulator value,
$ \gamma_q\!\ll\!\omega_q$, there may be an
intermediate regime in which the spin waves
are not yet absorbed by the electron-hole continuum,
even if the damping is increased
compared to the insulator.
To the extent this occurs,
the system can remain in the QC regime with
$z\!=\!1$ in a wide range of doping and temperatures
due to the Lorentz invariant terms in the
action.

In the insulator, the zero temperature energy gap
$ \Delta\!=\!\hbar c/\xi$
for the spin-1 excitations in a quantum disordered (QD) regime
is, again, related to the Lorentz invariance at $T\!=\!0$ \cite{CHN}.
Hence, as long as the
Lorentz invariant terms in the action are still
important and the correlation length saturates,
the low frequency ($ \omega<\Delta$) spectral weight
could be suppressed even in a metal. We suggest that
the {\em spin gap} phenomenon
in YBa$_2$Cu$_3$O$_{6.63}$ and YBa$_2$Cu$_4$O$_8$,
characterized by a sharp increase in $T_1T$
and decrease of the bulk susceptibility, is related to this
suppression, and this phase corresponds to a QD, $z\!=\!1$, regime.

At larger hole densities, the spin waves will be fully damped by the
electron-hole continuum. In this regime, the self-consistent
renormalization (SCR)
approach developed by Moriya {\em et al.} \cite{SCR}
(see also Ref.\cite{Millis:z=2})
and the phenomenological theory of Millis {\em et al.}
\cite{Millis:Monien:Pines},
for YBa$_2$Cu$_3$O$_7$ may be expected to apply. The
dynamical exponent $z\!=\!2$, while $\eta\!=\!0$;
in contrast with the previous case, the mean field exponent
$z\!=\!2$ is not fixed by a symmetry, but rather follows from the
scaling analysis of Ref.\cite{SCR,Millis:z=2}.
One obtains
$ T_1T/T_{\rm 2G}^2\!\sim\!\bar{\omega}\xi^2\!\sim\!\xi^{2-z}
\!\simeq\!\mbox{const}$,
while $ T_1T/T_{\rm 2G}\!\sim\!\xi^{-1}$.
At high temperatures,
the energy scale $\bar{\omega}\!\sim\!T$, so that
$\xi\!\sim\!\bar{\omega}^{-1/z}\!\sim\!T^{-1/z}\!\sim\!T^{-1/2}$,
in which case $ T_1T/T_{\rm 2G}\!\sim\!T^{1/2} $, and, separately,
$1/T_1\!\sim\!T\xi^z\!\simeq\!\mbox{const} $,
$ 1/T_{\rm 2G}\!\sim\!\xi\!\sim\!T^{-1/2}$.
We emphasize that $1/T_1\!\simeq\!\mbox{const}$ at high temperatures
is predicted for both $z\!=\!1$ and $z\!=\!2$ regimes, while
predictions for $1/T_{\rm 2G}$ are different.
Finally, at still larger hole densities, the short range
AF correlations between spins will tend to disappear; in this limit,
$ \xi\!\alt\!a$ is independent of temperature, and
one recovers the Korringa law,
$ 1/T_1\!\sim\!T $,
while $ 1/T_{\rm 2G}\!\simeq\!\mbox{const}$.
This regime corresponds to a normal metal, in which any remaining
antiferromagnetic correlations can be
described by a temperature independent $F^a({\bf p,p}')$.

\bigskip
\noindent
{\bf 3. La$_{2-x}$Sr$_x$CuO$_4$}
\bigskip

We first consider
La$_{2-x}$Sr$_x$CuO$_4$.
The insulator, La$_2$CuO$_4$, is well described
by the 2D Heisenberg model with the nearest-neighbor
exchange coupling $J\simeq 1500K$, except near or below
$T_N\!\sim\!300\mbox{ K}$ induced by weak interplanar
coupling \cite{reviews}.
A nearly temperature independent $1/T_1$
is observed in the insulating La$_2$CuO$_4$ above
$650\mbox{K}$ \cite{ImaiT1}, as expected in the QC,
$z\!=\!1$, regime \cite{Chubukov:Sachdev}.
The absolute value of $1/T_1\simeq 2700\,\mbox{sec}^{-1}$
at high temperatures \cite{ImaiT1} is in very good agreement
with both $1/N$ expansion
\cite{Chubukov:Sachdev} and finite cluster \cite{Sokol:Gagliano:Bacci}
calculations for the Heisenberg model.
Further, the ratio $ T_1T/T_{\rm 2G} $ measured in the insulator
\cite{ImaiT1,ImaiT2}
is nearly temperature independent in the broad range
$ 450\,\mbox{K}\!<\!T\!<\!900\,\mbox{K}$ (Fig.\ref{T1T}).
This is again what one would expect in the QC, $z\!=\!1$, regime;
we note that this behavior holds even
in the region below $650\,\mbox{K}$,
where $1/T_1$ and $T/T_{\rm 2G}$ separately
deviate from constant values, apparently because
$T_1T/T_{\rm 2G}$ is insensitive to the magnitude
of $\chi({\bf q},\omega)$.

A finite cluster calculation in the $S\!=\!1/2$ 2D Heisenberg model
by E. Gagliano, S. Bacci, and one of the authors (A.S.)
(Ref.\cite{Sokol:Gagliano:Bacci} and this work),
with no adjustable parameters
used (hyperfine and exchange couplings were determined
from other experiments, see Ref.\cite{Sokol:Gagliano:Bacci}),
indeed yields a nearly temperature independent
$ T_1T/T_{\rm 2G} \simeq  4.3 \cdot 10^3 \mbox{K}^{-1} $
for $T\!>\!J/2\!\simeq\!750$K (Fig.\ref{T1T}),
in excellent agreement with the experimental
result, $ 4.5 \cdot 10^3 \mbox{K} $ \cite{ImaiT1}.
The systematic error of the finite cluster calculation,
arising from the periodic boundary conditions as well as
spin diffusion contribution to $1/T_1$,
is estimated in Ref.\cite{Sokol:Gagliano:Bacci} as
$10\!-\!15$\%.

An especially striking feature of the $1/T_1$ data \cite{ImaiT1}
is the nearly doping independent absolute value
of $1/T_1$ in the high temperature limit (Fig.\ref{T1T}).
This result shows that
not only universal, but also nonuniversal scaling constants
are not strongly renormalized in the doping range
$x\!=\!0\!-\!0.15$,
i.e.\ up to the {\em optimal} concentration for the superconductivity.
Since it would be rather unrealistic to assume that
exactly the same value can be obtained in different pictures
of magnetism for low- and high-doped La-based materials,
we suggest that the high temperature magnetic behavior over this
entire doping range has the same physical origin as that found
for the insulating state, which implies $z\!=\!1$.
As the doping increases, Imai, Slichter, and collaborators
\cite{ImaiT1} find that the range of temperatures where
$T_1T$ is linear in temperature
stretches towards lower temperatures,
from $ 650\, \mbox{K} $ for $x\!=\!0$
down to $125\,\mbox{K}$ for $x\!=\!0.15$
(see Fig.\ref{T1T}).
This is the behavior expected if doping leads
to a decrease in $\rho_s$, thus extending
the QC region \cite{Chubukov:Sachdev}.
We thus conclude that
La$_{1.85}$Sr$_{0.15}$CuO$_4$ is in the
quantum-critical, $z\!=\!1$, regime
for $T\!\agt\!125\,\mbox{K}$.

As it is evident from Fig.\ref{T1T} (inset),
for $T\!\alt\!125$K, $T_1T$
begins to depart from its linear in $T$ behavior, exhibiting
an upturn for $T\!\approx\!60\!-\!70\,\mbox{K}$.
We attribute this effect to the suppression of the low frequency
spectral weight in the quantum disordered, $z\!=\!1$, regime
\cite{Millis:Monien}. At lower doping values,
experiment shows that the low temperature
phase is not a superconductor but rather a spin glass,
in agreement with the scaling analysis \cite{Sachdev:Ye},
while for the lowest doping, the low temperature phase
is the antiferromagnetic Neel state.

\newpage
\bigskip
\noindent
{\bf 4. YBa$_2$Cu$_3$O$_{6.63}$ and YBa$_2$Cu$_4$O$_8$}
\bigskip

YBa$_2$Cu$_3$O$_{6.63}$ and YBa$_2$Cu$_4$O$_8$
have quite similar properties.
The product $T_1T$ measured in YBa$_2$Cu$_3$O$_{6.63}$
\cite{TakigawaT1,TakigawaT2},
is linear in temperature for $160\mbox{K}\!\alt\!T\!\alt\!300\mbox{K}$,
while it exhibits similar behavior
in YBa$_2$Cu$_4$O$_8$ for $170\mbox{K}\!\alt\!T\!\alt\!800\mbox{K}$.
Since a linear $T_1T$ is predicted in both quantum-critical
($z\!=\!1$)
and overdamped ($z\!=\!2$) regimes at high temperatures,
to distinguish between these regimes,
we turn to the $1/T_{\rm 2G}$ data on
YBa$_2$Cu$_3$O$_{6.63}$ \cite{TakigawaT2}, and plot
$ T_1T/T_{\rm 2G} $ and $ T_1T/T_{\rm 2G} $
as a function of temperature (Fig.\ref{ratios}).
In the range $200\,\mbox{K}\!<\!T\!<\!300\,\mbox{K}$,
$T_1T/T_{\rm 2G}$ is nearly constant, while
$T_1T/T_{\rm 2G}^2$ varies significantly,
in agreement with the prediction for $z\!=\!1$.
Were this material in the $z\!=\!2$ regime,
$T_1T/T_{\rm 2G}$ would increase as the temperature increases,
while $T_1T/T_{\rm 2G}^2$ would be constant.
We thus conclude that above $200$K,
YBa$_2$Cu$_3$O$_{6.63}$,
and the closely related YBa$_2$Cu$_4$O$_8$,
are in the QC, $z\!=\!1$, regime.
The increase in damping in the doped case, which
enhances $1/T_1$ with respect to $1/T_{\rm 2G}$,
may explain the smaller (compared to the insulating
La$_2$CuO$_4$, Fig.\ref{T1T})
saturation value of $T_1T/T_{\rm 2G}$ in those compounds.
Our scenario may seem to contradict the
Raman studies in YBa$_2$Cu$_3$O$_{6+x}$,
because two-magnon Raman scattering
is not observed for doping
above O$_{6.4}$ \cite{Cooper}. However,
this contradiction is illusory, since
the decrease of intensity in the two-magnon Raman scattering
is primarily due to the loss of the charge-transfer states
rather than any change in the short range magnetic correlations
\cite{Cooper,Klein}.

For temperatures below $150\,\mbox{K}$, $1/T_1$ sharply drops down as
the temperature decreases, while $1/T_{\rm 2G}$
\cite{TakigawaT1,TakigawaT2} saturates.
As was the case for La$_{1.85}$Sr$_{0.15}$CuO$_4$,
we argue this suppression of the low frequency spectral weight
({\em spin gap}) reflects a crossover
to the quantum disordered, $z\!=\!1$, regime.
Since in this regime the magnitude of the
gap is inversely proportional to the correlation length, which is
smaller in YBa$_2$Cu$_3$O$_{6.63}$
and YBa$_2$Cu$_4$O$_8$ than in
La$_{1.85}$Sr$_{0.15}$CuO$_4$, the onset temperature
of the quantum disordered (spin gap) regime is larger and
the crossover to it is more pronounced in the former two materials.
Thus, unlike the scenario proposed by Millis and Monien
\cite{Millis:Monien}, we argue that for all three materials
the physical origin of the spin gap is the same.

\bigskip
\noindent
{\bf 5. YBa$_2$Cu$_3$O$_7$}
\bigskip

For nearly stoichiometric YBa$_2$Cu$_3$O$_{6.9}$, $ T_1T/T_{\rm 2G}$
\cite{Imai7} increases as the temperature increases,
while $ T_1T/T_{\rm 2G}^2$ is nearly constant
(Fig.\ref{ratios}), in agreement with the scaling prediction
for the {\em overdamped}, $z\!=\!2$, regime \cite{SCR}.
The departure from the Korringa law $1/T_1\!\sim\! T $ and
large copper-to-oxygen ratio of the relaxation rates shows that
the antiferromagnetic enhancement is still quite substantial.
In the overdamped regime, the spin wave branch is either destroyed,
or due to the small correlation length
has appreciable spectral weight only for energies much larger than
the maximal temperature at which this compound is chemically stable.
Therefore, for experimentally accessible temperatures
no departure from the the overdamped regime is observed.

\bigskip
\noindent
{\bf 6. Conclusion}
\bigskip

We have shown that the nuclear relaxation data
in a broad range of doping levels, which includes
both metallic and insulating materials,
possesses universal features
characteristic of the quantum critical regime of a clean
antiferromagnetic insulator
with the dynamical exponent $z\!=\!1$.
This universality suggests that well beyond the
metal-insulator transition, the spin excitation spectrum of the
insulator is not yet
destroyed by the electron-hole background,
in which case one expects a
two-component dynamics for a broad range of doping levels.
While a two-component {\em dynamics}
may arise in a one-component as well as a two-component
{\em microscopic model}, we call attention to an
explicit example which leads to this kind of dynamics directly,
namely, a model of spins and fermions
with both spin-spin ($J$) and weak
spin-fermion ($\widetilde J$) exchange interaction
\cite{Schrieffer}. The robustness of the spin waves may be related to
either their weak coupling to quasiparticles, or to quasiparticle
Fermi surfaces which are not spanned by
the antiferromagnetic ordering vector ${\bf Q}$.

On the basis of the above analysis,
we suggest the unified magnetic phase diagram for the cuprate
superconductors shown in Fig.\ref{PD};
the proposed boundary between the QC and QD regimes is
determined from the nuclear relaxation data shown on Fig.\ref{T1T}.
We propose that as the hole doping increases,
the transition from the insulating to the
overdoped regime occurs in two stages. First, the system
becomes metallic; the damping of spin waves
increases somewhat
compared to its value in the insulator, but since the spin waves
are not destroyed by the electron-hole background,
the dynamical exponent is $z\!=\!1$
and the quantum critical regime persists over a wide range
of temperatures and doping levels.
Then, at substantially higher doping,
the dynamical exponent crosses over to $z\!=\!2$.
We further argue that the spin gap
phenomenon observed in the underdoped materials
reflects the same physics
as the formation of the gap
for spin excitations in the the quantum disordered, $z\!=\!1$,
phase of an insulator.
This scenario suggests that
in compounds where the spin gap is observed,
the temperature dependent bulk susceptibility should exhibit
a downturn near the crossover from the QC to QD regime,
while $T_1T/T_{\rm 2G}$ should be
temperature independent at higher temperatures.
We show, in a subsequent communication \cite{BPST},
that our scenario leads in a natural way to a unified
description of the results of nuclear relaxation,
magnetic susceptibility, and neutron scattering experiments.

\bigskip
\noindent
{\bf Acknowledgements}
\bigskip

We are grateful to M.\ Takigawa, T.\ Imai, and C.P.\ Slichter
for communicating their experimental data to us
in advance of publication.
We would like to thank V.\ Barzykin, G.\ Blumberg, A.V.\ Chubukov,
S.L.\ Cooper, E.\ Dagotto,
D.\ Frenkel, L.P.\ Gor'kov, M.V.\ Klein, A.J.\ Millis,
and D.\ Thelen for valuable discussions,
and T.\ Imai and C.P.\ Slichter for
numerous stimulating conversations on nuclear resonance phenomena.
This work has been supported by the NSF Grant DMR89-20538 through
the Materials Research Laboratory.

\figure{The proposed magnetic
phase diagram for the cuprate superconductors above $100$K.
\label{PD}}

\figure{Experimental data on
$T_1T$ and $T_1T/T_{\rm 2G}$: \
$\otimes$ La$_2$CuO$_4$ \cite{ImaiT1,ImaiT2}; \newline
$\nabla$ La$_{1.85}$Sr$_{0.15}$CuO$_4$ \cite{ImaiT1}; \
$\Delta$ La$_{1.85}$Sr$_{0.15}$CuO$_4$ \cite{Kitaoka}; \
$\bullet$ YBa$_2$Cu$_4$O$_8$ \cite{Machi}; \
$\circ$ YBa$_2$Cu$_3$O$_{6.63}$ \cite{TakigawaT1,TakigawaT2}; \
\rule{5pt}{5pt} YBa$_2$Cu$_3$O$_{6.9}$ \cite{Imai7}.
Also shown ($\Diamond$) are the results of numerical calculation
of $1/T_1$ \cite{Sokol:Gagliano:Bacci} and
$1/T_{\rm 2G}$ (\cite{Sokol:Gagliano:Bacci} and
this work) for the insulator.
The arrows indicate our proposed values for the crossover temperature
from the QC to QD regimes; the inset makes clear our choice for
La$_{1.85}$Sr$_{0.15}$CuO$_4$.
\label{T1T}}

\figure{Experimental data on
$T_1T/T_{\rm 2G}$ and $T_1T/T_{\rm 2G}^2$:
\ $\circ$ YBa$_2$Cu$_3$O$_{6.63}$
\cite{TakigawaT1,TakigawaT2};
\newline \rule{5pt}{5pt} YBa$_2$Cu$_3$O$_{6.9}$ \cite{Imai7}.
\label{ratios}}


\begin{references}

\bibitem{Imai7}
T. Imai, C.P. Slichter, A.P. Paulikas, and B. Veal,
Phys. Rev. B {\bf 47}, 9158 (1993).

\bibitem{ImaiT1}
T. Imai, C.P. Slichter,
K. Yoshimura, and K. Kosuge,
Phys. Rev. Lett. {\bf 70}, 1002 (1993).

\bibitem{TakigawaT2}
M. Takigawa, private communication.

\bibitem{ImaiT2}
T. Imai, C.P. Slichter, K. Yoshimura,
M. Katoh, and K. Kosuge, to be published.

\bibitem{Pennington:Slichter}
C.H. Pennington and C.P. Slichter,
Phys. Rev. Lett. {\bf 66}, 381 (1991);
see also
D. Thelen and D. Pines, preprint (1993).

\bibitem{Singh}
R.R.P. Singh, Phys. Rev. B {\bf 39}, 9760 (1989).

\bibitem{CHN}
S. Chakravarty, B.I. Halperin, and D.R. Nelson,
Phys. Rev. B {\bf 39}, 2344 (1989).

\bibitem{Sachdev:Ye}
S. Sachdev and J. Ye, Phys. Rev. Lett. {\bf 69}, 2411 (1992).

\bibitem{Chubukov:Sachdev}
A.V. Chubukov and S. Sachdev, preprint No.\ 9301027;
A.V.\ Chubukov, S.\ Sachdev, and J.\ Ye, preprint
No.\ 9304046 in cond-mat$@$babbage.sissa.it.

\bibitem{Makivic:Jarrell}
M. Makivi\'c and M. Jarrell, Phys. Rev. Lett. {\bf 68}, 1770 (1992).

\bibitem{Sokol:Gagliano:Bacci}
A. Sokol, E. Gagliano, and S. Bacci, Phys. Rev. B, June 1, 1993
(in press; see No.\ 9302013 in cond-mat$@$babbage.sissa.it);
and to be published.

\bibitem{Hayden}
S.M. Hayden, G. Aeppli, R. Osborn, A.D. Taylor, T.G. Perring,
S.-W. Cheong, and Z. Fisk, Phys. Rev. Lett. {\bf 67}, 3622 (1991).

\bibitem{Millis:z=2}
A.J. Millis, preprint (1993).

\bibitem{SCR}
T.\ Moriya, Y.\ Takahashi, and K.\ Ueda,
Phys. Rev. Lett. {\bf 59}, 2905 (1990), and references
therein.

\bibitem{Millis:Monien:Pines}
A.J. Millis, H. Monien, and D. Pines,
Phys. Rev. B {\bf 42}, 167 (1990).

\bibitem{reviews}
S. Chakravarty, in Proceedings of
{\em High Temperature Superconductivity}, edited by K.S. Bedell
{\em et al.} (Addison-Wesley, CA, 1990);
E. Manousakis, Rev. Mod. Phys. {\bf 63}, 1 (1991).

\bibitem{Millis:Monien}
A.J.\ Millis and H.\ Monien [Phys.\ Rev.\ Lett.\ {\bf 70}, 2810 (1993)]
have suggested that temperature dependent bulk susceptibility in
YBa$_2$Cu$_3$O$_{6.63}$ is due to the gap for spin excitations
induced by the interplanar coupling,
while in La$_{1.85}$Sr$_{0.15}$CuO$_4$
it is caused by the SDW fluctuations, in contrast with our
common mechanism for both Y- and La-based materials.
The spin gap regime has been discussed in a number of publications;
some of them are listed as Ref.7 of the above paper.

\bibitem{TakigawaT1}
M. Takigawa, A.P. Reyes, P.C. Hammel, J.D. Thompson,
R.H. Heffner, Z. Fisk, and K.C. Ott,
Phys. Rev. B {\bf 43}, 247 (1991).

\bibitem{Cooper}
S.L. Cooper, D. Reznik, A. Kotz, M.A. Karlow, R. Liu, M.V. Klein,
W.C. Lee, J. Giapintzakis, and D.M. Ginsberg,
Phys. Rev. B, {\bf 47}, 8233 (1993).

\bibitem{Klein}
We thank M.V.\ Klein for pointing this out to us.

\bibitem{Schrieffer}
J.R. Schrieffer, L. Lilly, and N.E. Bonesteel,
APS 1993 March Meeting abstract G23-1, and private
communication; see also L.P. Gor'kov and A. Sokol,
Physica C {\bf 159}, 329 (1989).

\bibitem{BPST}
V. Barzykin, D. Pines, A. Sokol, and D. Thelen, to be published.

\bibitem{Kitaoka}
Y. Kitaoka, S. Ohsugi, K. Ishida, and K. Asayama,
Physica C {\bf 189}, 189 (1990).

\bibitem{Machi}
T. Machi, I. Tomeno, T. Miyatake, N. Koshizuka, S. Tanaka,
T. Imai, and H. Yasuoka,
Physica C {\bf 173}, 32 (1991).

\end{references}
\end{document}